\documentclass[twocolumn,showpacs,jcp,superscriptaddress]{revtex4}

\usepackage{graphicx}
\usepackage[usenames]{color}
\usepackage{amssymb}
\usepackage{amsmath}
\usepackage{amsfonts}
\usepackage{mathrsfs}

\renewcommand{\epsilon}{\varepsilon}
\newcommand{\figurewidth}{0.44\textwidth}

\begin{document}
\title{Translocation Dynamics with Attractive Nanopore-Polymer Interactions}

\author{Kaifu Luo}
\altaffiliation[]{
Author to whom the correspondence should be addressed}
\email{luokaifu@gmail.com}
\affiliation{Department of Engineering Physics, Helsinki University of Technology,
P.O. Box 1100, FIN-02015 TKK, Espoo, Finland}
\author{Tapio Ala-Nissila}
\affiliation{Department of Engineering Physics, Helsinki University of Technology,
P.O. Box 1100, FIN-02015 TKK, Espoo, Finland}
\affiliation{Department of Physics, Box 1843, Brown University, Providence,
Rhode Island 02912-1843, USA}
\author{See-Chen Ying}
\affiliation{Department of Physics, Box 1843, Brown University, Providence,
Rhode Island 02912-1843, USA}
\author{Aniket Bhattacharya}
\affiliation{Department of Physics, University of Central Florida, Orlando,
Florida 32816-2385, USA}

\date{\today}
\begin{abstract}

Using Langevin dynamics simulations, we investigate the influence of
polymer-pore interactions on the dynamics of biopolymer translocation
through nanopores. We find that an attractive interaction can significantly 
change the translocation dynamics. This can be understood by
examining the three components of the total translocation time
$\tau \approx \tau_1+\tau_2+\tau_3$ corresponding to the initial filling
of the pore, transfer of polymer from the \textit{cis} side to the
\textit{trans} side, and emptying of the pore, respectively.
We find that the dynamics for the last process of emptying of the pore 
changes from non-activated to activated in nature as the strength of 
the attractive interaction increases, 
and $\tau_3$ becomes the dominant contribution to the total translocation 
time for strong attraction. This leads to a new dependence of $\tau$
as a function of driving force and chain length.
Our results are in good agreement with recent experimental findings, and 
provide a possible explanation for the different scaling behavior observed 
in solid state nanopores {\it vs.} that for the natural $\alpha$-hemolysin channel.

\end{abstract}

\maketitle
\section{Introduction}

The controlled transport of polymer molecules through a nanopore
has received increasing attention due to its importance in
biological systems and its potentially revolutionary
technological applications~\cite{Kasianowicz,Meller03}.
There is a flurry of experimental
~\cite{Akeson,Meller00,Meller01,Meller02,Meller07,Bashir,Sauer,Mathe,Henrickson,
Li01,Li03,Li05,Keyser1,Keyser2,Dekker,Trepagnier,Storm03,Storm052,Storm05}
and theoretical
~\cite{Storm05,Simon,Sung,Park,diMarzio,Muthukumar99,MuthuKumar03,Lubensky,Kafri,Slonkina,
Ambj,Metzler,Ambj2,Ambj3,Baumg,Chuang,Kantor,Panja2,Panja,Dubbeldam1,Dubbeldam2,Milchev,
Luo1,Luo2,Luo3,Luo4,Huopaniemi1,Huopaniemi2,Luo5,Luo6,LuoComment,Slater,Chern,Loebl,Randel,
Lansac,Kong,Farkas,Tian,Lu,Liao,LuoMB,Zandi,Tsuchiya,Matysiak,Bhattacharya}
studies devoted to this subject.
In an important experiment, Kasianowicz \textit{et al.}~\cite{Kasianowicz}
demonstrated that an electric field can drive single-stranded DNA and RNA
molecules through the water-filled $\alpha$-hemolysin channel and that the
passage of each molecule is signaled by a blockade in the channel current.
These observations can be used to directly characterize the polymer length.
Similar experiments have been done recently using solid state nanopores with
more precisely controlled dimensions~\cite{Li01,Li03,Li05,Trepagnier,Keyser1,Keyser2,
Dekker,Storm03,Storm052,Storm05}.
Currently, extensive effort is being taken to unravel the
dependence of the translocation time $\tau$ on the system parameters
such as the polymer chain length
$N$~\cite{Meller01,Meller02,Storm05,Sung,Muthukumar99,MuthuKumar03,Lubensky,Chuang,Kantor,Panja2,
Panja,Dubbeldam1,Dubbeldam2,Milchev,Luo1,Luo2,Luo3,Luo4,Huopaniemi1,Huopaniemi2,
Tian,Matysiak},
pore length $L$ and pore width $W$~\cite{Luo1}, driving force $F$
~\cite{Meller01,Meller02,Henrickson,Sauer,Kantor,Luo2,Huopaniemi1,Tian,Matysiak},
sequence and secondary
structure~\cite{Akeson,Meller00,Meller02,Luo3,Luo4},
and polymer-pore
interactions~\cite{Meller00,Meller02,Lubensky,Tian,Luo4,Luo5,Luo6,LuoMB}.

Meller \textit{et al.}~\cite{Meller00,Meller02} have shown how several
different DNA polymers can be identified by a unique pattern in an ``event diagram''.
The event diagrams are plots of translocation duration versus blockade current for an
ensemble of events.
Patterns for a given polymer can be characterized uniquely by the blockade
current, the translocation time and its distribution. Because each type of
polynucleotide gives rise to specific values of these three parameters, DNA
molecules which differ from each other only by sequence can be distinguished.
At room temperature striking differences were found for the translocation time
distributions of polydeoxyadenylic acid (poly(dA)$_{100}$) and polydeoxycytidylic
acid (poly(dC)$_{100}$) DNA molecules. The translocation time of poly(dA) is found
to be much longer, and its distribution is wider with a longer tail compared with
the corresponding data for poly(dC). Moreover, the differences between the translocation
behavior are accentuated at lower temperature.
The origin of the different behavior was attributed to stronger attractive
interaction of poly(dA) with the pore.
Recently Krasilnikov \textit{et al}.~\cite{Krasilnikov} have investigated the
dynamics of single neutral poly (ethylene glycol) (PEG) molecules in the $\alpha$-hemolysin
channel in the limit of a strong attractive polymer-pore attraction.
The result for the residence time in the channel shows a novel non-monotonic behavior
as a function of the molecular weight.
The other experimental data that point to the possible essential role of the monomer-pore
interaction concerns the different conflicting values of scaling exponents of $\tau$ with
$N$ and with the applied voltage as reported in recent experiments.
A linear dependence $\tau \sim N$ was observed for polymer translocation
through $\alpha$-hemolysin channel~\cite{Kasianowicz,Meller01}, while
another experiment reported that $\tau \sim N^{1.27} \approx N^{2\nu}$ 
for a synthetic nanopore~\cite{Storm05}, where $\nu$ is
the Flory exponent~\cite{de Gennes,Doi}.
As to the dependence of the translocation time on the applied voltage for
$\alpha$-hemolysin channel, an inverse linear behavior~\cite{Kasianowicz} is observed
for polyuridylic acid (poly(U)) while an inverse quadratic dependence~\cite{Meller01}
is found for polydeoxyadenylic acid (poly(dA)). One possible explanation for all these
conflicting data is that the polymer-pore interaction depends crucially on the details
of the pore structure ($\alpha$-hemolysin channel vs synthetic nanopore) in addtion to
being base pair specific.

To date, most of the theoretical studies of the translocation of biopolymers through
nanopre are based on models in which the wall of the pore only plays a passive role
in confining the polymer to the inside of the pore. 
There are only a few theoretical studies of such interaction effects.
Based on a Smoluchowski equation with a phenomenological microscopic potential to
describe the polymer-pore interactions, Lubensky and Nelson~\cite{Lubensky} captured the
main ingredients of the translocation process. However, when comparing with experiments, their
model is not sufficient. Numerically, Tian and Smith~\cite{Tian} found that attraction
facilitates the translocation process by shortening the translocation time, which contradicts
experimental findings~\cite{Meller00,Meller02}.
In a recent letter~\cite{Luo4}, we used Langevin dynamics (LD) simulations to investigate the
influence of polymer-pore interactions on translocation.
We found that with increasing attraction, the histogram for the translocation time $\tau$ shows
a transition from Gaussian distribution to a long-tailed distribution corresponding to thermal
activation over a free energy barrier.
The $N$ dependence of the entropic force leads to both the translocation time and the residence
time in the pore showing a non-monotonic behavior as a function of $N$ for short chains in the
strong attraction limit. These results are in good agreement with the above experimental
data~\cite{Meller00,Meller02,Krasilnikov}.

In the present work, we further show that strong
polymer-pore interactions can directly affect the {\it effective} scaling exponents of $\tau$
both with $N$ and  with the applied voltage, which provides a possible explanation for the different
experimental findings~\cite{Kasianowicz,Meller01,Storm05} on these physical quantities.
We provide a microscopic understanding of how strong polymer-pore interaction influences the
translocation dynamics. This is done through analyzing the three quantities $\tau_1$, $\tau_2$ and
$\tau_3$ corresponding to initial filling of the pore, transfer of the polymer from the \textit{cis}
side to the \textit{trans} side, and finally emptying of the pore, respectively. We find that the
final process of emptying the pore $\tau_3$ involves an activation barrier and completely dominates
the translocation time in the strong attractive interaction limit.
This leads to a strong dependence of the {\it effective} scaling exponents associated with
the translocation time {\it both} on the strength of the attraction {\it and}
the driving force. In addtion, we examine the waiting time and residence time distributions.
These quantities are related to the translocation time but the waiting time provides more detailed
information about the translocation dynamics, while the residence time is the more relevant quantity
for direct comparison with with the experimental observation.


\section{Model and methods} \label{chap-model}

In our numerical simulations, the polymer chains are
modeled as bead-spring chains of Lennard-Jones (LJ) particles with
the Finite Extension Nonlinear Elastic (FENE) potential. Excluded
volume interaction between monomers is modeled by a short range
repulsive LJ potential: $U_{LJ} (r)=4\epsilon
[{(\frac{\sigma}{r})}^{12}-{(\frac{\sigma} {r})}^6]+\epsilon$ for
$r\le 2^{1/6}\sigma$ and 0 for $r>2^{1/6}\sigma$. Here, $\sigma$ is
the diameter of a monomer, and $\epsilon$ is the depth of the
potential. The connectivity between neighboring monomers is modeled
as a FENE spring with $U_{FENE}
(r)=-\frac{1}{2}kR_0^2\ln(1-r^2/R_0^2)$, where $r$ is the distance
between consecutive monomers, $k$ is the spring constant and $R_0$
is the maximum allowed separation between connected monomers.

\begin{figure}
  \includegraphics*[width=\figurewidth]{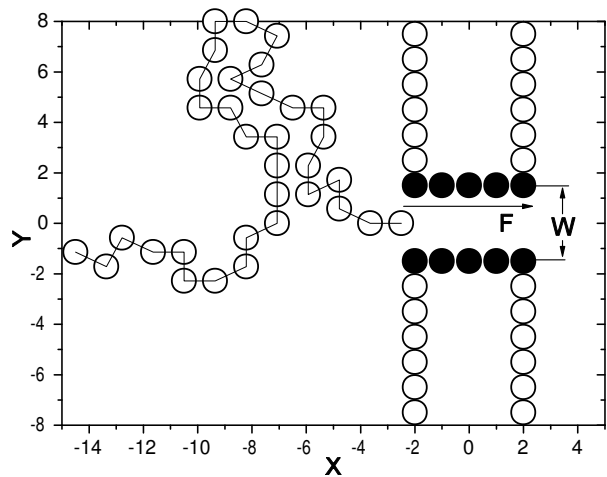}
\caption{ A schematic representation of the system. The pore length
$L=5 $ and the pore width $W=3$ (see text for the units).
        }
 \label{Fig1}
\end{figure}

We consider a 2D geometry as shown in Fig. \ref{Fig1}, where the
wall in the $y$ direction is described as stationary particles
within a distance $\sigma$ from each other. The pore of length $L$
and width $W$ in the center of the wall is composed of stationary
black particles.
Between all monomer-wall particle pairs, there exist the same short
range repulsive LJ interaction as described above.
The pore-monomer interaction is modeled by a LJ potential with
a cutoff of $2.5\sigma$ and interaction strength $\epsilon_{pm}$.
This interaction can be either attractive or repulsive depending on
the position of the monomer from the pore particles.
In the Langevin dynamics simulation, each monomer is subjected to
conservative, frictional, and random forces, respectively,
with~\cite{Allen} $m{\bf \ddot
{r}}_i =-{\bf \nabla}({U}_{LJ}+{U}_{FENE})+{\bf F}_{\textrm{ext}}
-\xi {\bf v}_i + {\bf F}_i^R$, where $m$ is the monomer's mass,
$\xi$ is the friction coefficient, ${\bf v}_i$ is the monomer's
velocity, and ${\bf F}_i^R$ is the random force which satisfies the
fluctuation-dissipation theorem.
The external force is expressed as ${\bf F}_{\textrm{ext}}=F\hat{x}$,
where $F$ is the external force strength exerted on the monomers in the pore,
and $\hat{x}$ is a unit vector in the direction along the pore axis.

In the present work, we use the LJ parameters $\epsilon$ and
$\sigma$ and the monomer mass $m$ to fix the energy, length and
mass scales respectively. Time scale is then given by
$t_{LJ}=(m\sigma^2/\epsilon)^{1/2}$. The dimensionless parameters in
our simulations are $R_0=2$, $k=7$, $\xi=0.7$ and $k_{B}T=1.2$
unless otherwise stated.
For the pore, we set $L=5$ unless otherwise stated. A choice of
$W=3$ ensures that the polymer encounters an attractive force inside
the pore.
The driving force $F$ is set between $0.5$ and $2.0$, which
correspond to the range of voltages used in the
experiments~\cite{Kasianowicz,Meller01}.
The Langevin equation is integrated in time by a method described by
Ermak and Buckholz~\cite{Ermak} in 2D.
Initially, the first monomer of the chain is placed in the entrance of
the pore, while the remaining monomers are under thermal collisions described
by the Langevin thermostat to obtain an equilibrium configuration.
Typically, we average our data over 2000 independent runs.

\section{Results and discussion} \label{chap-results}

\subsection{Translocation time, waiting time and residence time}

\begin{figure}
\includegraphics*[width=\figurewidth]{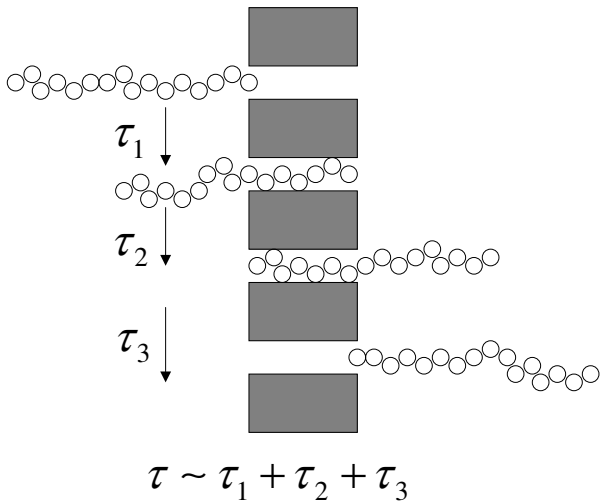}
\caption{Three components of the translocation process.
        }
\label{Fig2}
\end{figure}

The translocation time is defined as the time interval between the entrance of the
first segment into the pore and the exit of the last segment. We can break down the
translocation process into three components, as shown in Fig. 2.
The total translocation time $\tau$  can be written as a sum of three contributions
$\tau \approx \tau_1 + \tau_2 + \tau_3$, where $\tau_1$, $\tau_2$ and $\tau_3$
correspond to initial filling of the pore, transfer of the polymer from
the \textit{cis} side to the \textit{trans} side, and finally emptying of the pore,
respectively.
To shed light on the detailed translocation process, we examine the number of
translocated monomers $n_{trans}$ as a function of the time for a typical
successful translocation event for $N=128$, and two values of the monomer attractive interaction strength.
The value $\epsilon_{pm}=1.0$ corresponds to a weak interaction whereas $\epsilon_{pm}=3.0$ corresponds to
the strong attraction limit.
Here, $n_{trans}=0$ before the first monomer exits
the pore and $n_{trans}=N$ after the last monomer has threaded through the pore.
As shown in Fig. 3, under the weak driving force $F=0.5$, $\tau_1$ is not
sensitive to the attraction strength and $\tau_1 \ll \tau_2$.
$\tau_2$ for the strong attraction with $\epsilon_{pm}=3.0$ is roughly twice as
that for the weak attraction with $\epsilon_{pm}=1.0$.
However, $\tau_3$ depends strongly on the attraction strength. For
$\epsilon_{pm}=1.0$, $\tau_3 \ll \tau_2$ and is basically negligible for the pore length $L=5$.
For the strong attraction limit with  $\epsilon_{pm}=3.0$, the situation is totally different with
$\tau_3$ more than an order of magnitude larger than $\tau_2$, completely dominating the total
contribution to the translocation time.
From Fig. 3, it can be seen that the number of translocated monomers  oscillates around
$n_{trans} \approx 122$, which corresponds to the beginning of the last
stage of the translocation process, namely the emptying of the pore.
This is due to the activated nature of the translocation process with a free energy difference of
$\Delta \widetilde{F}=L(\epsilon_{pm}-F/2-f(N))$ between the final and the
initial state. The term $f(N)$ here accounts for the entropic
driving force which should kick in at larger values of $N$, and
eventually saturate for very long polymers. This leads to the long oscillation time of
the last few monomers with repeated forward and backward motions.
The final emptying of the pore corresponds to a rare crossing of the barrier.

\begin{figure}
\includegraphics*[width=\figurewidth]{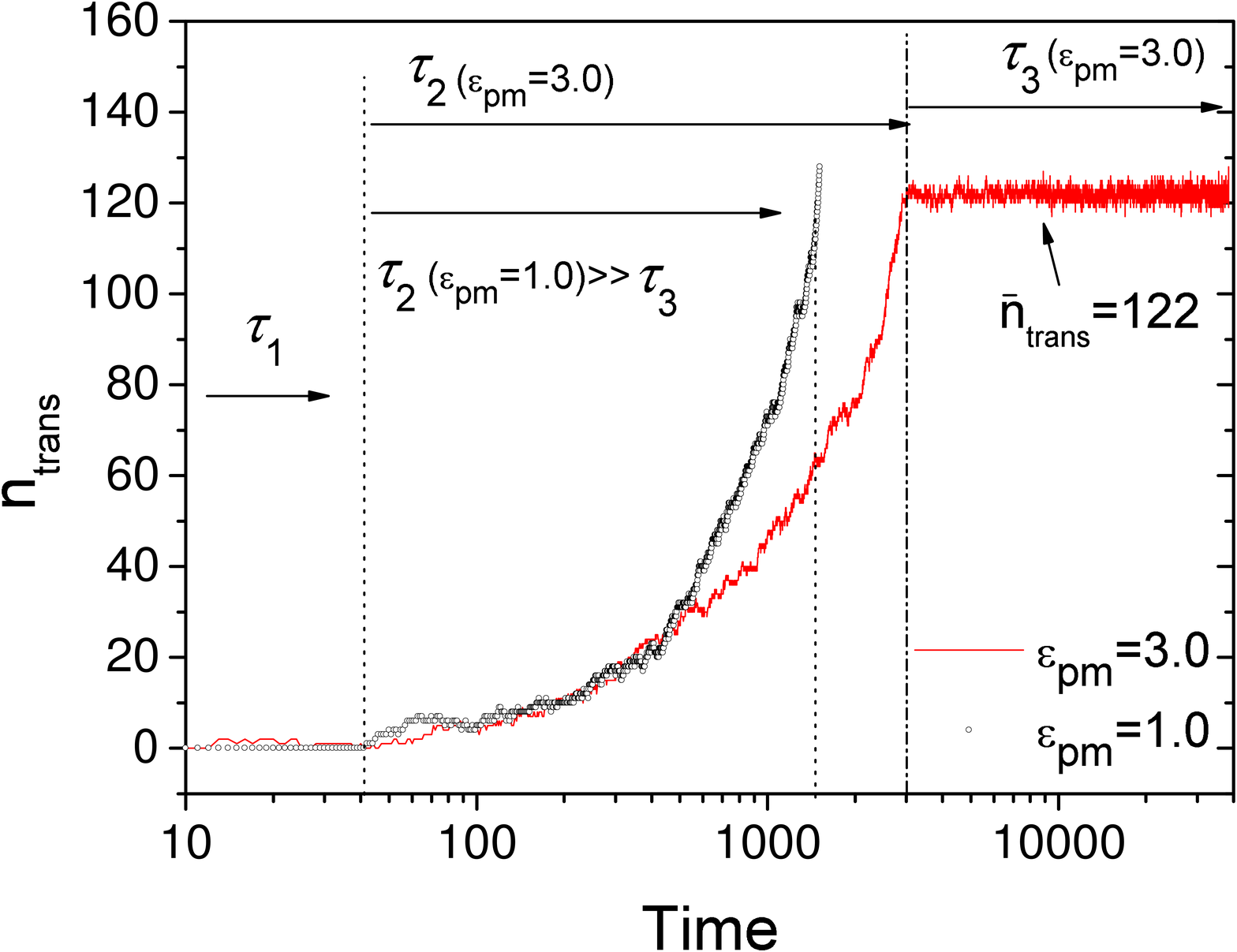}
\caption{
Number of translocated monomers $n_{trans}$ as a function of the time for
$\epsilon_{pm}=1.0$ and $\epsilon_{pm}=3.0$ under the driving force $F=0.5$.
For both strong and weak attraction strengths, $\tau_1 \ll \tau$.
For weak attraction strength $\epsilon_{pm}=1.0$, we find $\tau_3 \ll \tau_2$
and thus $\tau \approx \tau_2$.
        }
\label{Fig3}
\end{figure}

\begin{figure}
\includegraphics*[width=\figurewidth]{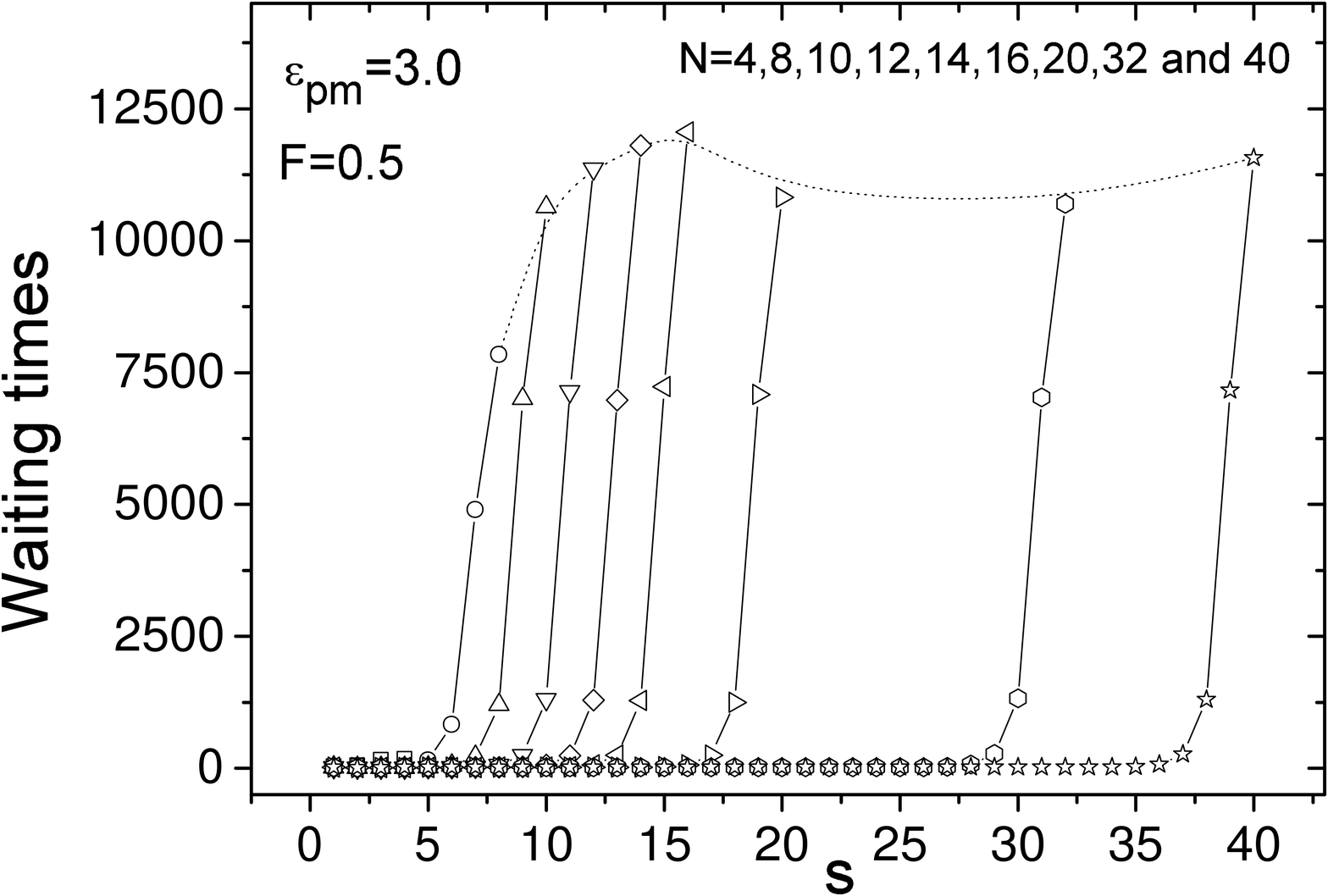}
\caption{
Waiting time of different chain lengths for
$\epsilon_{pm}=3.0$ and $F=0.5$.
        }
\label{Fig4}
\end{figure}

To provide more microscopic details of the translocation process, we investigate the
waiting time distribution for different chain lengths $N$ in the strong attraction limit.
The waiting time of monomer $s$ is defined as the average time between
the events that monomer $s$ and monomer $s+1$ exit the pore. 
In our previous work~\cite{Luo2,Huopaniemi1} for pure repulsive monomer-pore 
interactions, we found that the waiting time depends strongly 
on the monomer positions in the chain. For relatively short polymers, the monomers in 
the middle of the polymer need the longest time to exit the pore. 
Here, the waiting time of different chain lengths for $\epsilon_{pm}=3.0$ and $F=0.5$ are 
shown in Fig. 4.
It can be seen that it takes much longer time for last three monomers to exit the pore, 
which is completely different from that for for pure repulsive monomer-pore interactions.
This behavior correlates with the oscillation of the last monomers as shown in Fig. 3.
Here we should mention that due to the entropic factor $f(N)$ in
the barrier the waiting time for these last few monomers
actually decreases in the range $N\approx 14-32$ before saturating
and even increasing slightly with further increase of $N$. 

\begin{figure}
\includegraphics*[width=\figurewidth]{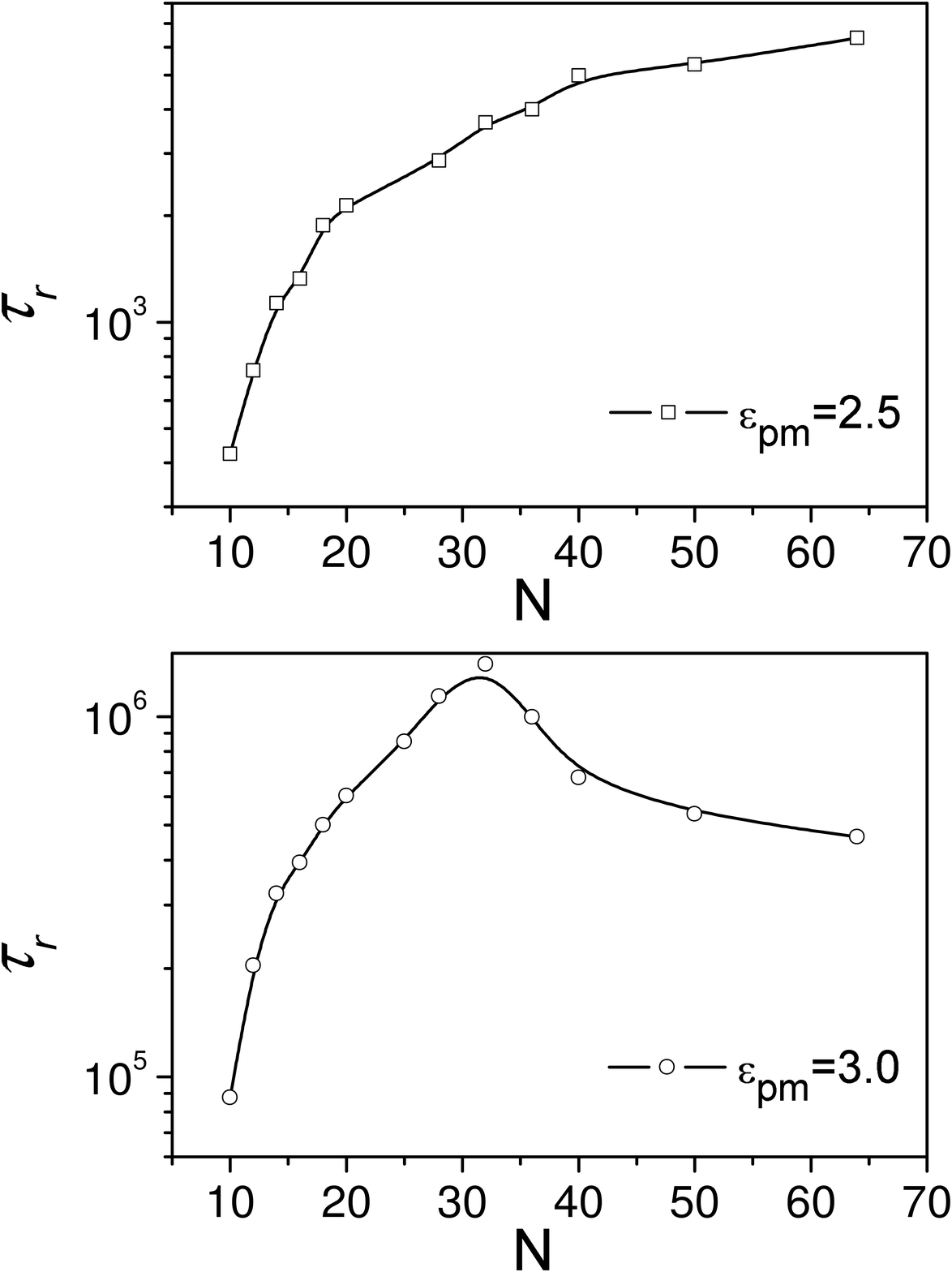}
\caption{
Residence time $\tau_r$ as a function of the chain length for
$\epsilon_{pm}=3.0$ and $\epsilon_{pm}=2.5$ under the driving force $F=0$.}
\label{Fig5}
\end{figure}

Under zero and low driving forces, the translocation probability
is very small in the sense that many translocation events, once started do not
finish all the way. Instead, the polymer returns and exits to the cis side again.
This means that the $\tau_1$ process of filling the pore do not get completed and
the real translocation process corresponding to $\tau_2$ and $\tau_3$ never even
get started. We define an additional residence time $\tau_r$ as the weighted average
of the translocation time for the completed events and the return time for the events
that start and return via the cis side. The significance of this quantity is that it
corresponds to the experimentally measured average blockage time of the polymer in the
nanopore which does not distinguish return events from the completed translocated events.
For zero or low driving force ($F<0.5$), the residence time is almost completely
dominated by return events.
We have calculated the residence time $\tau_r$ for $\epsilon_{pm}=2.5$ and 3 in Fig. 5.
As shown in Ref.~\onlinecite{Luo4}, in the strong attraction case with $\epsilon_{pm}=3.0$,
the $N$ dependence of the residence time here is non-monotonic.
This result of $\tau_r$ is in good agreement with experimental data of
Krasilnikov~\textit{et al}.~\cite{Krasilnikov} where the residence time of a neutral PEG
molecule in $\alpha$-hemolysin pore was measured.
Here, we further show that for $\epsilon_{pm}=2.5$, $\tau_r$ increases with increasing $N$.
It indicates that the strong attraction plays an essential role in the observed non-monotonic
behavior.

\begin{figure}
\includegraphics*[width=\figurewidth]{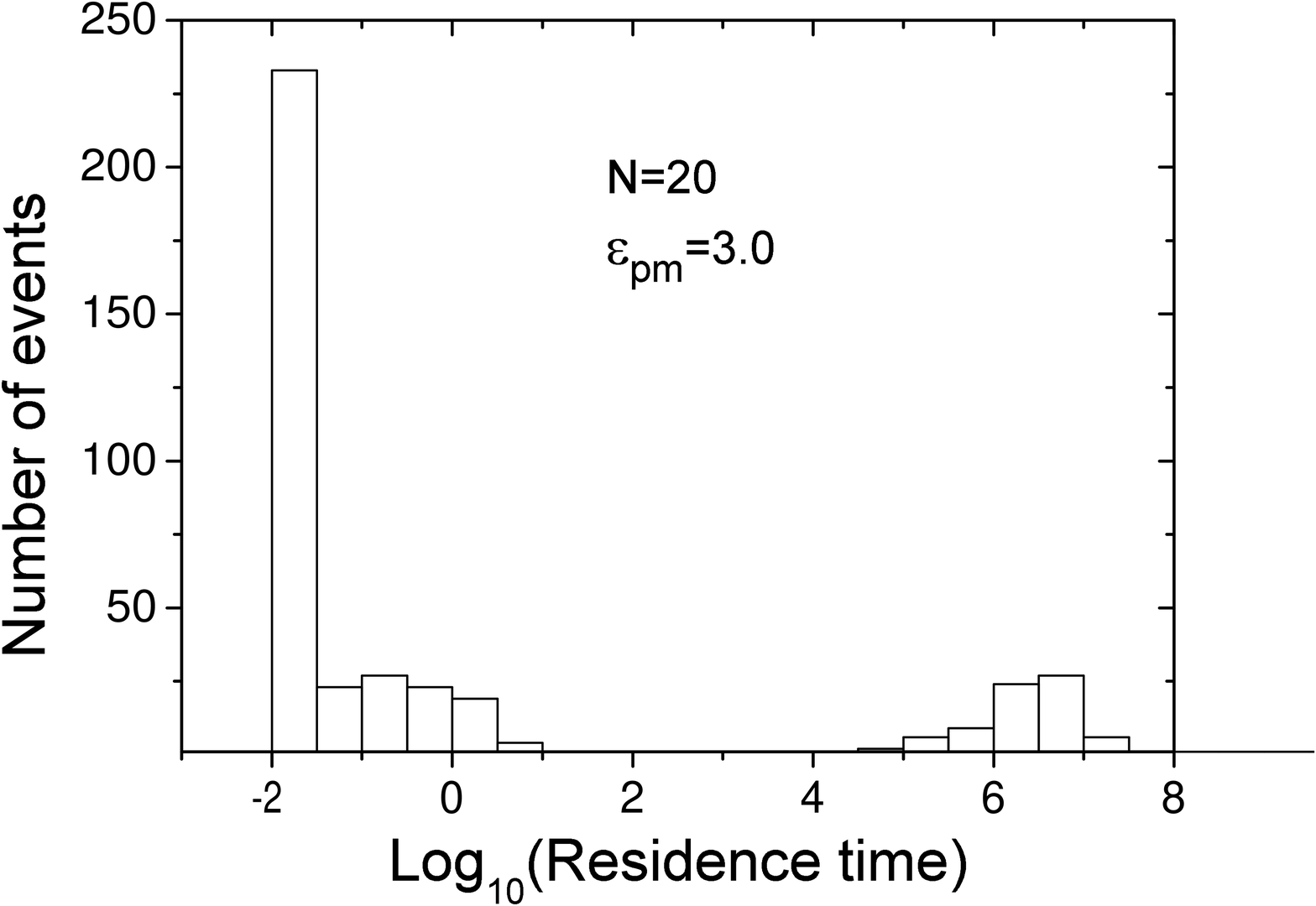}
\caption{
Distribution of the residence time for
$\epsilon_{pm}=3.0$ and $F=0$. The chain length $N=20$. }
\label{Fig6}
\end{figure}

For $\epsilon_{pm}=3.0$, the distribution of $\tau_r$ is shown in Fig. 6.
One obvious feature is the existence of two groups. The first group with shorter $\tau_r$
corresponds to the events where one end of the chain accesses the pore, and then quickly
returns back. For the second group with longer $\tau_r$, the
residence time is still about 99.8$\%$ due to return events for $\epsilon_{pm}=3.0$.
In the strong attraction limit, once the attractive force reaches its maximum when
the pore is fully filled by monomers, it takes a very long time for the polymer to return back
due to frequent backward and forward events.

\subsection{Dependence of translocation time on various parameters}
\subsubsection{Translocation time as a function of temperature}

Fig. 7 shows the translocation time $\tau$ as a function of the temperature for
different attraction strengths.
For the whole examined range of temperatures, $\tau$ decreases very slightly with increasing
temperature for a weak attractive strength of $\epsilon_{pm}=1.0$.
However, for the strong attractive strength $\epsilon_{pm}=3.0$, with increasing
temperature $\tau$ first rapidly decreases and then approaches saturation at higher
temperatures.
At higher temperatures, the differences between translocation times for weak and
strong attractive strengths become very small.
This temperature dependence of translocation time is in good
agreement with experiments~\cite{Meller00}.

\begin{figure}
  \includegraphics*[width=\figurewidth]{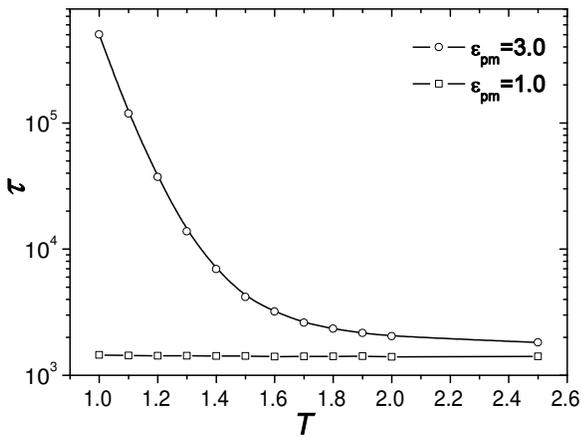}
\caption{Translocation time as a function of the temperature for both strong and weak attraction
strengths ($\epsilon_{pm}=3.0$ and $1$, respectively)
under the driving force $F=0.5$. The chain length $N=128$.}
 \label{Fig7}
\end{figure}

\subsubsection{Translocation time as a function of the driving force}

\begin{figure}
  \includegraphics*[width=\figurewidth]{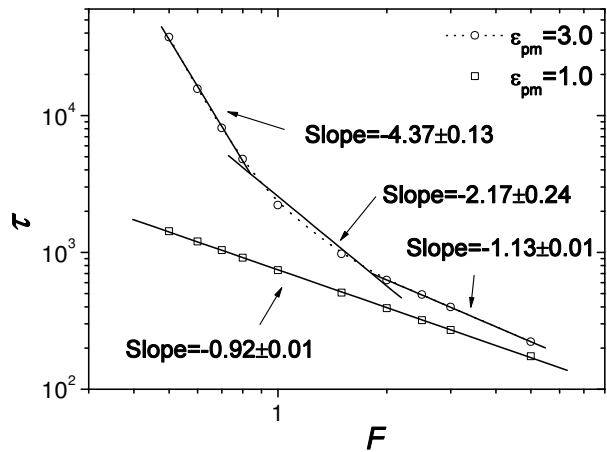}
\caption{Translocation time as a function of the driving forces for both strong
and weak attraction strengths, $\epsilon_{pm} =3$ and 1.
The chain length $N=128$.
}
\label{Fig8}
\end{figure}

In the weak attraction ({\it i.e.} non-activated) region, the overall $\tau$ is determined mainly by
$\tau_2$ and its dependence on the driving force scales as $F^{-1}$. This simple scaling behavior can
be understood by considering the steady state of motion of the polymer through the nanopore. The average
velcity is determined by balancing the frictional damping force (proprtional to the velocity) with the
external driving force. This leads to an average velocity proportinal to the driving force $F$, and hence
a translocation time $\tau \sim F^{-1}$.
In Fig. 8 we show the dependence of the translocation time $\tau$ on the driving force.
It can be seen that in the weak interaction limit for $\epsilon_{pm}=1.0$ the data are very close to the
linear scaling behavior $\tau \sim F^{-1}$ as predicted.
For strong attractive inteaction with $\epsilon_{pm}=4.0$, the situation is more complicated.
For weak driving forces ($F \le 2$), one is in the the activated region where the inverse of the
translocation time obeys an Arrhenius form. However, the driving force $F$ affects both the activation
barrier and the prefactor, leading to a complicated dependence of $\tau$ on the driving force that does
not have a simple power law scaling form as seen in Fig. 8 for the $\epsilon_{pm}=3.0$ result.
Insistence on fitting the data with a power law scaling form will lead to an effective scaling exponent
that changes with the value of the driving force. 
Finally, beyond a critical force, the activation barrier disppears and one should obtain asymptotically
the $\tau \sim F^{-1}$ behavior just as in the weak interaction case.
This whole scenario is very similar to the sliding friction of an adsorbed
layer under an external driving force~\cite{See-chen}.

The above theoretical considerations lead to a possible explanation of recent apparently conflicting experimental
data. Polyuridylic acid (poly(U)) has a weak interaction with the pore, and it is not surprising that an inverse
linear dependence of the translocation time on applied voltage
was observed in experiments on the translocation of poly(U)~\cite{Kasianowicz}.
However, Poly(dA) has much stronger interaction with the pore compared with
poly(U). Thus it should be in the strong interaction  activated region with a larger {\it effective} scaling exponent.
Indeed, an inverse quadratic dependence of the translocation time on applied voltage had been experimentally
observed for poly(dA)~\cite{Meller01}. In view of our theoretical considerations, it would be desirable to have
measurement over a larger range of the applied voltage to see the predicted change of effective scaling exponent.

\subsubsection{Translocation time as a function of chain length }

Previously, we have established that for pure repulsive polymer-pore interactions,
the dependence of the translocation time on the length of the polymer scales as
$\tau\sim N^{2\nu}$ for $N<200$ and crosses over to a new scaling regime $\tau \sim N^{1+\nu}$
for larger values of $N$~\cite{Luo2,Huopaniemi1,Luo3}. 
In the presence of weak interaction between the monomer and the pore,
the qualitative dependence on the length of the polymer remains the same.
For stronger attractive strength $\epsilon_{pm}=3.0$, the scaling exponent
of $\tau$ with $N$ for $64 \le N \le 400$ becomes strongly dependent on
the driving force, with no indication of crossover behavior as shown
in Fig. 9. We find $\tau\sim N^{1.32}$ for $F=2.0$, which is close to
$\tau\sim N^{2\nu}$ with the Flory exponent $\nu=0.75$ in 2D~\cite{de Gennes,Doi},
and $\tau\sim N^{0.97}$ for $F=1.0$.
The novel dependence on the length of polymer is due to the change from the non-activated 
regime for weak attractive or pure repulsive interaction to an activated regime for strong 
attractive interaction.

\begin{figure}
  \includegraphics*[width=\figurewidth]{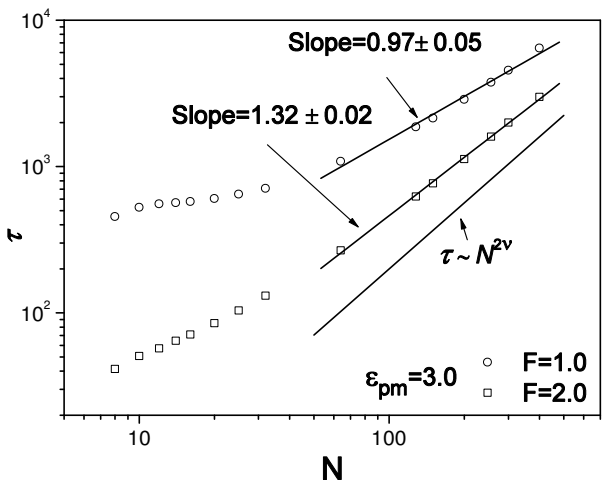}
\caption{
    Translocation time as a function of the chain length for
$\epsilon_{pm}=3.0$ under $F=1.0$ and $F=2.0$, respectively.
        }
 \label{Fig9}
\end{figure}

Experimentally, a linear dependence $\tau \sim N$  was observed in
experiments~\cite{Kasianowicz,Meller01} for polymer translocation
through $\alpha $-hemolysin channel, in contrast to the $\tau \sim N^{2\nu}$
scaling observed for polymer translocation through the solid-state
nanopore~\cite{Storm05}. This difference can be understood in light of our
present results concerning the influence of the different polymer-pore interaction
on the length dependence of the translocation time.
For a synthetic pore, there is at most a very weak attractive interaction
between the polymer and the pore and one expects the scaling behavior
$\tau\sim N^{2\nu}$ to hold for $N \le 200$.
However, a stronger attractive interaction is expected to exist between the
different bases and the $\alpha$-hemolysin channel. 
For the models studied in this work, it changes the scaling behavior 
from $\tau \sim N^{2\nu}$ to $\tau \sim N$. This provides a possible 
explanation for the difference of the experimental observations in the 
different nanopores~\cite{Kasianowicz,Meller01,Storm05}.

Under a strong attractive force with $\epsilon_{pm}=3.0$ and a weak
driving force $F=0.5$, the translocation time $\tau$ has a
qualitatively different dependence on $N$ as compared with the pure
repulsive or weak attractive pore interaction. Here we should mention that for
$F=0.5$ we cannot access $N>128$ as the translocation time becomes too long to be feasible for
numerical comuputation. As shown earlier in Ref.~\onlinecite{Luo4} and here in
Fig. 10, the translocation time displays
novel non-monotonic behavior with a rapid increase to a maximum at $N \sim 14$,
followed by a decrease for $14<N<32$ and an increases again for
$N>32$. The eventual increase in the large $N$ limit is due to the
$\tau_2$ contribution for longer chains.
The observed non-monotonic behavior is also reflected qualitatively in the
waiting time distribution as shown in Fig. 4.
As shown in Fig. 10, with decreasing an attractive force to $\epsilon_{pm}=2.5$,
this non-monotonic behavior vanishes.

\begin{figure}
\includegraphics*[width=\figurewidth]{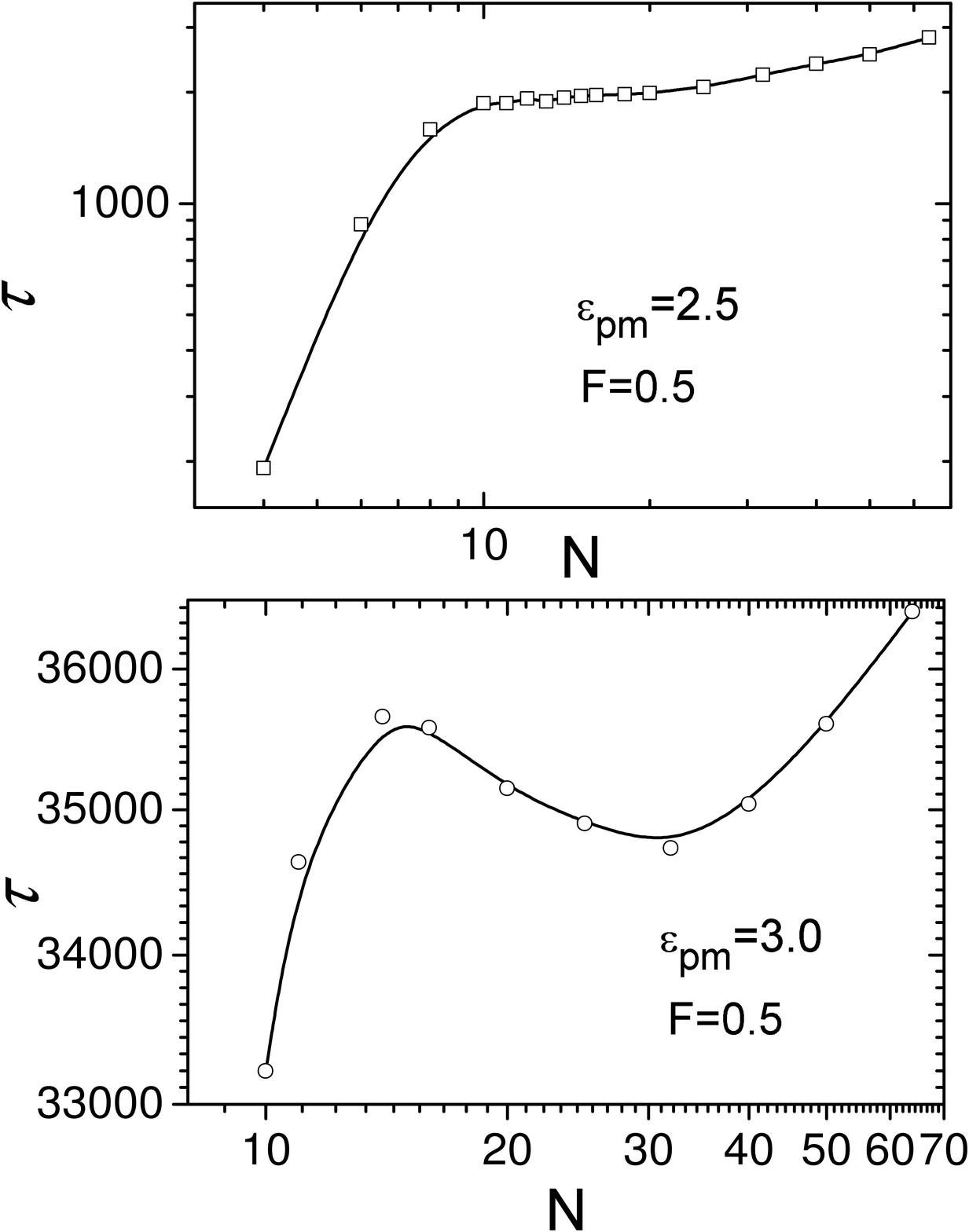}
\caption{
Translocation time $\tau$ as a function of the chain length for
$\epsilon_{pm}=3.0$ and $\epsilon_{pm}=2.5$ under the driving force $F=0.5$.
        }
\label{Fig10}
\end{figure}

To understand the microscopic origin, in Fig. 11 we show $\tau_1+\tau_2$ as a 
function of the chain length for different attraction strengths under the driving 
force $F=0.5$. For $32 \le N \le 200$, $\tau_1+\tau_2 \sim N^{2\nu}$ is observed, 
irrespective of attraction strengths.
This indicates that the novel non-monotonic behavior shown in Fig.10 in the strong 
interaction limit is again due to the pore-emptying process corresponding to $\tau_3$ 
dominating the translocation time in strong interaction limit.

\begin{figure}
  \includegraphics*[width=\figurewidth]{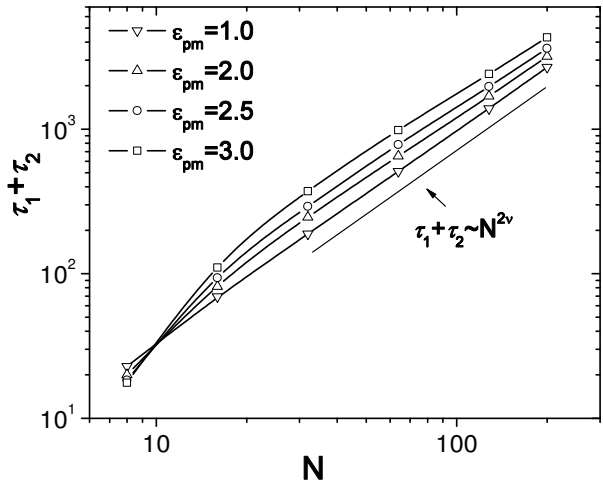}
\caption{
    $\tau_1+\tau_2$ as a function of the chain length for different
$\epsilon_{pm}$ under the driving force $F=0.5$.
        }
 \label{Fig11}
\end{figure}

\section{Conclusions} \label{chap-conclusions}

In this work, we have studied the dependence of the translocation
time on the temperature, attraction strength, driving force and the chain length.
To analyze the influence of the attractive interaction in more detail,
we have considered the three components of the translocation time
$\tau \approx \tau_1+\tau_2+\tau_3$, which were examined as a function of the attraction
strength. Here $\tau_1$, $\tau_2$ and $\tau_3$ correspond to
initial filling of the pore, transfer of polymer from the \textit{cis} side to
the \textit{trans} side, and emptying of the pore, respectively.
We find that $\tau_1 \ll \tau_2$ for both weak
and strong attraction strengths, for $N$ in the typical range used in the experiments.
However, $\tau_3$ is sensitive to the presence of an attractive interaction and changes from
a value much less than $\tau_2$ for weak interactions to the dominant contribution to the
overall translocation time due to the rare activated event nature of the final emptying of
the pore. This leads to a drastic change of the translocation dynamics and various scaling
exponents as a function of the strength of the attractive monomer pore interactions.
Our theoretical results are  in good agreement with recent
experimental data \cite{Meller00,Meller02,Krasilnikov}. They also provide a possible explanation
for the difference of the scaling behaviors with regard to the driving  force and the length of
polymers observed using different types of nanopores\cite{Kasianowicz,Meller01,Storm05}.

\begin{acknowledgments}
This work has been supported in part by The Academy of Finland
through its Center of Excellence (COMP) and TransPoly Consortium grants.
\end{acknowledgments}


\begin{thebibliography}{8}

\bibitem{Kasianowicz} J. J. Kasianowicz, E. Brandin, D. Branton and D. W. Deamer, \textit{Proc. Natl. Acad. Sci. U.S.A.} {\bf 93}, 13770 (1996).
\bibitem{Meller03} A. Meller, \textit{J. Phys.: Condens. Matter} {\bf 15}, R581 (2003).
\bibitem{Akeson} M. Akeson, D. Branton, J. J. Kasianowicz, E. Brandin, and D. W. Deamer, \textit{Biophys. J.} {\bf 77}, 3227 (1999).
\bibitem{Meller00} A. Meller, L. Nivon, E. Brandin, J. A. Golovchenko, and D. Branton, \textit{Proc. Natl. Acad. Sci. U.S.A.} {\bf 97}, 1079 (2000).
\bibitem{Meller01} A. Meller, L. Nivon, and D. Branton, \textit{Phys. Rev. Lett.} {\bf 86}, 3435 (2001).
\bibitem{Meller02} A. Meller and D. Branton, \textit{Electrophoresis} {\bf 23}, 2583 (2002).
\bibitem{Meller07} M. Wanunu and A. Meller, \textit{Nano Lett.} {\bf 7}, 1580 (2007).
\bibitem{Bashir} S. M. Iqbal, D. Akin, and R. Bashir, \textit{Nat. Nanotech.} {\bf 2}, 243 (2007).
\bibitem{Sauer} A. F. Sauer-Budge, J. A. Nyamwanda, D. K. Lubensky, and D. Branton, \textit{Phys. Rev. Lett.} {\bf 90}, 238101 (2003).
\bibitem{Mathe} J. Mathe, H. Visram, V. Viasnoff, Y. Rabin, and A. Meller, \textit{Biophys. J.} {\bf 87}, 3205 (2004).
\bibitem{Henrickson} S. E. Henrickson, M. Misakian, B. Robertson, and J. J. Kasianowicz, \textit{Phys. Rev. Lett.} {\bf 85}, 3057 (2000).
\bibitem{Li01} J. L. Li, D. Stein, C. McMullan, D. Branton, M. J. Aziz, and J. A. Golovchenko, \textit{Nature} (London) {\bf 412}, 166 (2001).
\bibitem{Li03} J. L. Li, M. Gershow, D. Stein, E. Brandin, and J. A. Golovchenko, \textit{Nat. Mater.} {\bf 2}, 611 (2003).
\bibitem{Li05} D. Fologea, J. Uplinger, B. Thomas, D. S. McNabb, and J. L. Li, \textit{Nano Lett.} {\bf 5}, 1734 (2005).
\bibitem{Keyser1} U. F. Keyser, J. B. M. Koelman, S. van Dorp, D. Krapf, R. M. M. Smeets, S. G. Lemay, N. H. Dekker, and C. Dekker, \textit{Nat. Phys.} {\bf 2}, 473 (2006).
\bibitem{Keyser2} U. F. Keyser, J. van der Does, C. Dekker, and N. H. Dekker, \textit{Rev. Sci. Instr.}, {\bf 77}, 105105 (2006).
\bibitem{Dekker}  C. Dekker, \textit{Nat. Nanotech.} {\bf 2}, 209 (2007).
\bibitem{Trepagnier} E. H. Trepagnier, A. Radenovic, D. Sivak, P. Geissler, and J. Liphardt, \textit{Nano Lett.} {\bf 7}, 2824 (2007).
\bibitem{Storm03} A. J. Storm, J. H. Chen, X. S. Ling, H. W. Zandbergen, and C. Dekker, \textit{Nat. Mater.} {\bf 2}, 537 (2003).
\bibitem{Storm052} A. J. Storm, J. Chen, H. Zandbergen, and C. Dekker, \textit{Phys. Rev. E} {\bf 71}, 051903 (2005).
\bibitem{Storm05} A. J. Storm, C. Storm, J. Chen, H. Zandbergen, J. -F. Joanny and C. Dekker, \textit{Nano Lett.} {\bf 5}, 1193 (2005).
\bibitem{Simon} S. M. Simon, C. S. Peskin, and G. F. Oster, \textit{Proc. Natl. Acad. Sci. U.S.A.} {\bf 89}, 3770 (1992).
\bibitem{Sung}   W. Sung and P. J. Park, \textit{Phys. Rev. Lett.} {\bf 77}, 783 (1996).
\bibitem{Park} P. J. Park and W. Sung, \textit{J. Chem. Phys.}  {\bf 108}, 3013 (1998).
\bibitem{diMarzio} E. A. diMarzio and A. L. Mandell, \textit{J. Chem. Phys.}  {\bf 107}, 5510 (1997).
\bibitem{Muthukumar99}   M. Muthukumar, \textit{J. Chem. Phys.} {\bf 111}, 10371 (1999).
\bibitem{MuthuKumar03} M. Muthukumar, \textit{J. Chem. Phys.}  {\bf 118}, 5174 (2003).
\bibitem{Kong} C. Y. Kong and M. Muthukumar, \textit{Electrophoresis} {\bf 23}, 2697 (2002);
               C. Y. Kong and M. Muthukumar, \textit{J. Chem. Phys.} {\bf 120}, 3460 (2004);
               C. Y. Kong and M. Muthukumar, \textit{J. Am. Chem. Soc.} {\bf 127}, 18252 (2005);
               M. Muthukumar and C. Y. Kong, \textit{Proc. Natl. Acad. Sci. U.S.A.} {\bf 103}, 5273 (2006);
               C. Forrey and Muthukumar, \textit{J. Chem. Phys.} {\bf 127}, 015102 (2007);
               C. T. A. Wong and M. Muthukumar, \textit{J. Chem. Phys.} {\bf 128}, 154903 (2008).
\bibitem{Lubensky} D. K. Lubensky and D. R. Nelson, \textit{Biophys. J.} {\bf 77}, 1824 (1999).
\bibitem{Kafri} Y. Kafri, D. K. Lubensky, and D. R. Nelson, \textit{Biophys. J.} {\bf 86}, 3373 (2004).
\bibitem{Slonkina} E. Slonkina and A. B. Kolomeisky, \textit{J. Chem. Phys.}  {\bf 118}, 7112 (2003);
                   S. Kotsev and A. B. Kolomeisky, \textit{J. Chem. Phys.}  {\bf 125}, 084906 (2006);
                   A. Mohan, A. B. Kolomeisky and M. Pasquali, \textit{J. Chem. Phys.}  {\bf 128}, 125104 (2008).
\bibitem{Matysiak} S. Matysiak, A. Montesi, M. Pasquali, A. B. Kolomeisky, and C. Clementi, \textit{Phys. Rev. Lett.}  {\bf 96}, 118103 (2006).
\bibitem{Ambj} T. Ambjornsson, S. P. Apell, Z. Konkoli, E. A. DiMarzio, and J. J. Kasianowicz, \textit{J. Chem. Phys.}  {\bf 117}, 4063 (2002).
\bibitem{Metzler} R. Metzler and J. Klafter, \textit{Biophys. J.} {\bf 85}, 2776 (2003).
\bibitem{Ambj2} T. Ambjornsson and R. Metzler, \textit{Phys. Biol.} {\bf 1}, 19 (2004).
\bibitem{Ambj3} T. Ambjornsson, M. A. Lomholt, and R. Metzler, \textit{J. Phys.: Condens. Matter} {\bf 17}, S3945 (2005).
\bibitem{Baumg} A. Baumgartner and J. Skolnick, \textit{Phys. Rev. Lett.} {\bf 74}, 2142 (1995).
\bibitem{Chuang} J. Chuang, Y. Kantor and M. Kardar, \textit{Phys. Rev. E} {\bf 65}, 011802 (2002).
\bibitem{Kantor} Y. Kantor and M. Kardar,  \textit{Phys. Rev. E} {\bf 69}, 021806 (2004).
\bibitem{Panja2} J. K. Wolterink, G. T. Barkema, and D. Panja, \textit{Phys. Rev. Lett.} {\bf 96}, 208301 (2006);
                 D. Panja, G. T. Barkema, and R. C. Ball, \textit{J. Phys.: Condens. Matter} {\bf 19}, 432202 (2007);
                 D. Panja and G. T. Barkema, \textit{Biophys. J.} {\bf 94}, 1630 (2008).
\bibitem{Panja} D. Panja, G. T. Barkema, and R. C. Ball, \textit{J. Phys.: Condens. Matter} {\bf 20}, 075101 (2008);
                H. Vocks, D. Panja, G. T. Barkema, and R. C. Ball, \textit{J. Phys.: Condens. Matter} {\bf 20}, 095224 (2008).
\bibitem{Dubbeldam1} J. L. A. Dubbeldam, A. Milchev, V.G. Rostiashvili, and T.A. Vilgis, \textit{Phys. Rev. E} {\bf 76}, 010801(R) (2007).
\bibitem{Dubbeldam2} J. L. A. Dubbeldam, A. Milchev, V.G. Rostiashvili, and T.A. Vilgis, \textit{Europhys. Lett.} {\bf 79}, 18002 (2007).
\bibitem{Milchev} A. Milchev, K. Binder, and A. Bhattacharya, \textit{J. Chem. Phys.} {\bf 121}, 6042 (2004).
\bibitem{Luo1} K. F. Luo, T. Ala-Nissila, and S. C. Ying, \textit{J. Chem. Phys.} {\bf 124}, 034714 (2006).
\bibitem{Luo2} K. F. Luo, I. Huopaniemi, T. Ala-Nissila, and S. C. Ying, \textit{J. Chem. Phys.} {\bf 124}, 114704 (2006).
\bibitem{Huopaniemi1} I. Huopaniemi, K. F. Luo, T. Ala-Nissila, and S. C. Ying, \textit{J. Chem. Phys.} {\bf 125}, 124901 (2006).
\bibitem{Huopaniemi2} I. Huopaniemi, K. F. Luo, T. Ala-Nissila, and S. C. Ying, \textit{Phys. Rev. E} {\bf 75}, 061912 (2007).
\bibitem{Luo3} K. F. Luo, T. Ala-Nissila, S. C. Ying, and A. Bhattacharya, \textit{J. Chem. Phys.} {\bf 126}, 145101 (2007).
\bibitem{Luo4} K. F. Luo, T. Ala-Nissila, S. C. Ying, and A. Bhattacharya, \textit{Phys. Rev. Lett.} {\bf 99}, 148102 (2007).
\bibitem{Luo5} K. F. Luo, T. Ala-Nissila, S. C. Ying, and A. Bhattacharya, \textit{Phys. Rev. Lett.} {\bf 100}, 058101 (2008).
\bibitem{Luo6} K. F. Luo, T. Ala-Nissila, S. C. Ying, and A. Bhattacharya, to be punlished.
\bibitem{LuoComment} K. F. Luo, T. Ala-Nissila, S. C. Ying, P. Pomorski and M. kattunen, arXiv:0709.4615.
\bibitem{Slater} S. Guillouzic and G. W. Slater, \textit{Phys. Lett. A} {\bf 359}, 261 (2006);
                 M. G. Gauthier and G. W. Slater, \textit{Eur. Phys. J. E} {\bf 25}, 17 (2008);
                 M. G. Gauthier and G. W. Slater, \textit{J. Chem. Phys.} {\bf 128}, 065103 (2008).
\bibitem{Chern} S.-S. Chern, A. E. Cardenas, and R. D. Coalson, \textit{J. Chem. Phys.} {\bf 115}, 7772 (2001).
\bibitem{Loebl} H. C. Loebl, R. Randel, S. P. Goodwin, and C. C. Matthai, \textit{Phys. Rev. E} {\bf 67}, 041913 (2003).
\bibitem{Randel} R. Randel, H. C. Loebl, and C. C. Matthai, \textit{Macromol. Theory Simul.} {\bf 13}, 387 (2004).
\bibitem{Lansac} Y. Lansac, P. K. Maiti, and M. A. Glaser, \textit{Polymer} {\bf 45}, 3099 (2004).
\bibitem{Farkas} Z. Farkas, I. Derenyi, and T. Vicsek, \textit{J. Phys.: Condens. Matter} {\bf 15}, S1767 (2003).
\bibitem{Tian} P. Tian and G. D. Smith, \textit{J. Chem. Phys.}  {\bf 119}, 11475 (2003).
\bibitem{Lu} Y. D. He, H. J. Qian, Z. Y. Lu, and Z. S. Li, \textit{Polymer} {\bf 48}, 3601 (2007);
               Y. C. Chen, C. Wang, and M. Luo, \textit{J. Chem. Phys.} {\bf 127}, 044904 (2007);
               Y. J. Xie, H. Y. Yang, H. T. Yu, Q. W. Shi, X. P. Wang, and J. Chen, \textit{J. Chem. Phys.} {\bf 124}, 174906 (2006).
\bibitem{Liao} D. Wei, W. Yang, X. Jin, and Q. Liao, \textit{J. Chem. Phys.} {\bf 126}, 204901 (2007).
\bibitem{LuoMB} M. B. Luo, \textit{Polymer} {\bf 48}, 7679 (2007).
\bibitem{Zandi} R. Zandi, D. Reguera, J. Rudnick, and W. M. Gelbart, \textit{Proc. Natl. Acad. Sci. U.S.A.} {\bf 100}, 8649 (2003).
\bibitem{Tsuchiya} S. Tsuchiya and A. Matsuyama, \textit{Phys. Rev. E} {\bf 76}, 011801 (2007).
\bibitem{Bhattacharya} A. Bhattacharya {\it et al.}, unpublished (2008).
\bibitem{Krasilnikov} O. V. Krasilnikov, C. G. Rodrigues, and S. M. Bezrukov, \textit{Phys. Rev. Lett.} {\bf 97}, 018301 (2006).
\bibitem{de Gennes} P. G. de Gennes, \textit{Scaling Concepts in Polymer Physics} (Cornell University Press, Ithaca, NY, 1979).
\bibitem{Doi} M. Doi, and S. F. Edwards, \textit{The Theory of Polymer Dynamics} (Clarendon, Oxford, 1986).
\bibitem{Allen}M.P. Allen, D.J. Tildesley, \textit{Computer Simulation of Liquids} (Oxford University Press, 1987).
\bibitem{Ermak} D. L. Ermak and H. Buckholz, \textit{J. Comput. Phys.} {\bf 35}, 169 (1980).
\bibitem{See-chen} E. Granato and S. C. Ying, \textit{Phys. Rev. Lett.} {\bf 85}, 5368 (2000).


\end{thebibliography}
\end{document}